\documentclass[twocolumn,bibnotes,pre]{revtex4-1}
\usepackage[latin9]{inputenc}
\usepackage{amsmath}
\usepackage{amssymb}
\usepackage{graphicx}
\usepackage{esint}

\makeatletter
\@ifundefined{textcolor}{}
{%
 \definecolor{BLACK}{gray}{0}
 \definecolor{WHITE}{gray}{1}
 \definecolor{RED}{rgb}{1,0,0}
 \definecolor{GREEN}{rgb}{0,1,0}
 \definecolor{BLUE}{rgb}{0,0,1}
 \definecolor{CYAN}{cmyk}{1,0,0,0}
 \definecolor{MAGENTA}{cmyk}{0,1,0,0}
 \definecolor{YELLOW}{cmyk}{0,0,1,0}
}

\usepackage{epsfig}\usepackage{subfigure}
\usepackage{color}

\makeatother

\begin{document}

\title{Planar Sheets Meet Negative Curvature Liquid Interfaces}

\author{Zhenwei Yao}
\altaffiliation{Present address: Department of Materials Science and
Engineering, Northwestern University, Evanston, Illinois, 60208-3108, USA}
\author{Mark Bowick}
\email[e-mail:]{bowick@phy.syr.edu}
\author{Xu Ma}
\author{Rastko Sknepnek}
\affiliation{Department of Physics, Syracuse University, Syracuse, New York 13244-1130,
USA}

\begin{abstract}
If an inextensible thin sheet is adhered to a substrate with a negative
Gaussian curvature it will experience stress due to geometric
frustration. We analyze the consequences of such geometric frustration using
analytic arguments and numerical simulations. Both concentric wrinkles
and eye-like folds are shown to be compatible with negative curvatures.
Which pattern will be realized depends on the curvature of the substrate.
We discuss both types of folding patterns and determine the phase diagram
governing their appearance. 
\end{abstract}

\pacs{46.32.+x, 46.70.De, 68.08.-p}

\maketitle

Geometric frustration occurs in wrapping a spherical Mozartkugel (``Mozart
sphere'') with planar foil
\cite{sadoc2006geometrical,demaine2009wrapping,hure2011wrapping}.  The extra
circumference at the edge of the planar sheet compared to the spherical
substrate to which it is conforming gives rise to ridges, narrow deformed
regions that occupy a small fraction of the total available volume and along
which the energy is focused \cite{witten2007stress}.  Recently the frustration
of a thin circular elastic sheet of $\approx1\textrm{mm}$ size covering the cap
of a spherical droplet has been studied \cite{king2012elastic}.  At first fine
radial wrinkles appear at the edge of the sheet and then become unstable to
localized folds as the size of the droplet decreases.  Once expects the
situation to be quite different for a sheet conforming to a negative curvatures
surface {--} in this case the edge of the sheet is stretched tangentially as
opposed to being compressed on spherical geometry. A completely different
frustration pattern on the planar sheet is thus expected.  In this paper, we
study the wrinkle/fold structure in an elasto-capillary system on a flat sheet
conforming  to a negatively curved substrate. 

The system we treat has two parts: an inextensible thin elastic hydrophilic sheet
and a saddle-like fluid interface with negative curvature. The size
of the elastic sheet is taken to be much bigger than the elasto-capillary
length $\sqrt{\kappa/\sigma}$, so that surface tension $\sigma$
dominates over the bending rigidity $\kappa$ \cite{roman2010elasto}.
Note that a standard sheet of paper can be regarded as inextensible and be used
to demonstrate the bending of a thin elastic sheet \cite{marder2007crumpling}.
When such an inextensible elastic sheet is placed on a negative curvature
liquid interface the capillary force pulls the planar sheet into full
contact with the liquid interface.

\begin{figure}[h]
\begin{minipage}[t]{0.49\columnwidth}%
\begin{center}
\includegraphics[width=1.4in]{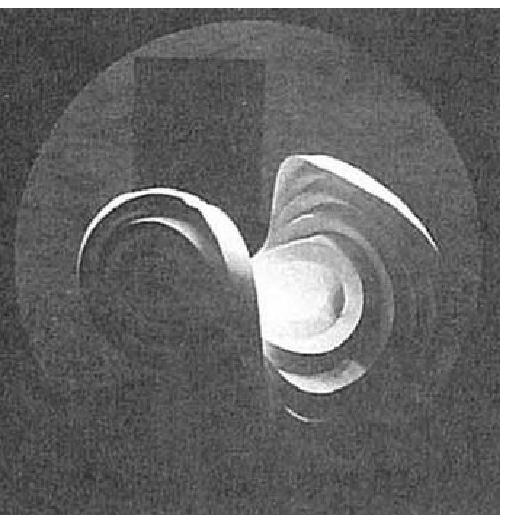} 
\par\end{center}%
\end{minipage}%
\begin{minipage}[t]{0.48\columnwidth}%
\begin{center}
\includegraphics[width=1.607in]{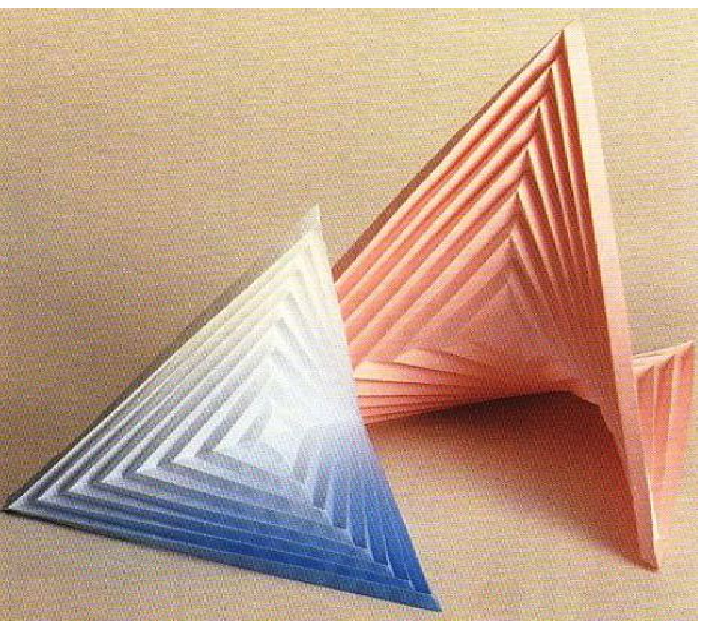}
\par\end{center}%
\end{minipage}

\caption{(color online) Wrinkles of alternating peaks and valleys on a piece
of paper redefine the metric, automatically bringing the flat paper
to a saddle-like shape. It is shown in the text that the effect of
concentric wrinkles is equivalent to inserting a wedge of some angle
in the sense of redefining the metric. (a) is excerpted from Ref.~\cite{wingler1978},
and (b) is from Ref.~\cite{jackson1989} where the method to make it
is described in detail. \label{fig:twophotos}}
\end{figure}

Complete contact between a planar sheet and a curved fluid substrate
by capillary forces introduces a wrinkle/fold pattern, which redefines
the metric of the planar sheet according to the curvature of the background
geometry. The modification of the metric leads to the change of shape.
This concept has been utilized to design responsive buckled surfaces
\cite{kim2012designing}. The curved surface endows its metric to
the flat sheet via their full contact. The inherited metric on the
flat sheet as well as its elasticity determines the wrinkle/fold structure.
Which wrinkle/fold patterns are compatible with negative curvatures?
The art of origami provides some inspiration. It is shown in origami
that regular concentric or square wrinkles with alternating peaks
and valleys on a piece of paper can induce a negative curvature, as
demonstrated in Fig.~\ref{fig:twophotos}.

In what follows we will prove that the effect of concentric wrinkles
is equivalent to inserting an angular wedge in the sense of redefining
the metric. The wavelength and amplitude of the concentric wrinkles
determine the angle of the wedge. The wrinkled sheet is parameterized
as $\vec{x}\left(r,\theta\right)=\left\{ r\cos\theta,r\sin\theta,a_{k}\cos\left(kr\right)\right\} $,
where $a_{k}$ is the amplitude of the wrinkles and $k$ is the wavenumber
$k=2\pi/\lambda$. The nonzero components of the metric tensor are
$g_{11}=1+x^{2}\sin^{2}\left(kr\right)$ and $g_{22}=r^{2}$, where
$x=a_{k}k$. The imposed wrinkles transform the original wrinkled
shell into a new surface denoted by $\Sigma$ that we take coincident
with the curved substrate. On this new surface the metric is redefined
such that the geodesic distance (denoted as $r$) from the center
of the disk to the first peak is $r_{0}=\lambda$. The corresponding
real distance on the original sheet is $l_{0}=\int_{0}^{\lambda}\sqrt{g_{11}}dr=\frac{1}{k}E\left(x\right)$,
where $E\left(x\right)=\int_{0}^{2\pi}dy\sqrt{1+x^{2}\sin^{2}y}$.
As $a_{k}\rightarrow0$, $l_{0}\rightarrow\lambda$, as expected.
Due to the inextensibility of the paper model, the mapping from the
originally flat sheet to a wrinkled shape is isometric and thus length-preserving.
The perimeter of the circle with radius $r$ in the new surface $\Sigma$
is thus $C\left(r\right)=2\pi l\left(r\right)\equiv2\pi r+r\alpha$,
where $l(r)=l_{0}kr/\left(2\pi\right)$. Here $\alpha$ is interpreted
as the angle of the inserted wedge: 
\begin{equation}
\alpha=2\pi\left(\frac{l_{0}}{\lambda}-1\right)>0\label{eq:alpha} \ .
\end{equation}
The positive sign of $\alpha$ indicates that concentric wrinkles
are equivalent to \textit{inserting} a wedge in redefining the metric.
The inserted wedge buckles a flat disk to a saddle-like shape with
negative curvature in three-dimensional Euclidean space \cite{audoly2010elasticity},
as do concentric wrinkles. These two ways of introducing negative
curvature - either inserting material or imposing concentric wrinkles
- are related via the expression for $C(r)$. The first method changes
the perimeter without changing the radius, while it is the opposite
for the second method. By expanding the expression for $l_{0}$ in
terms of small $x$, one finds $C(r)=2\pi r+\frac{\pi}{3}k^{2}x^{2}r^{3}+{\cal O}\left(x\right)$.
By inserting this into $K_{G}=\lim_{r\rightarrow0}3\left[2\pi r-C\left(r\right)\right]/\left(\pi r^{3}\right)$
\cite{struik1988lectures}, the curvature at the center of the surface
$\Sigma$ is found to be
\begin{equation}
K_{G}=-a_{k}^{2}k^{4}.\label{eq:KGwrinkle}
\end{equation}
Eq. \eqref{eq:KGwrinkle} shows that increasing the amplitude or the
frequency of the concentric wrinkles results in greater curvature
of the surface, with greater sensitivity to the frequency. The curvature
of the background geometry determines the amplitude and wavelength
of the concentric wrinkles according to the product $a_{k}^{2}k^{4}$.

It is worth mentioning that by increasing (``growing'') the radius
of a flat disk while keeping the perimeter invariant, we get a positive
curvature surface, which may buckle to various patterns depending
on its elasticity. This elasticity paradigm has been used to explain the phyllotactic
patterns of Fibonacci-like sequences on plants \cite{shipman2004phyllotactic}.
Wrinkles of concentric squares can buckle a square piece of paper
to a beautiful hyperbolic parabola with negative curvature as shown
in Fig.~\ref{fig:twophotos}(b). Prescribed metrics via the design
of wrinkles can even transform a flat piece of paper into a rich variety of
structures, including the DNA double helix \cite{kasahara2003extreme}.

\begin{figure}[h]
\centering{} \includegraphics[width=3in]{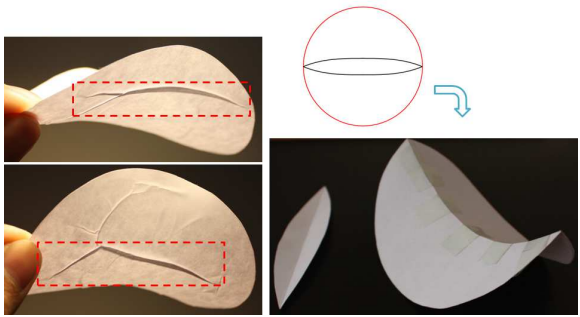} \caption{(color online) A branch-like fold pattern appears on a flat paper
disk by wrapping it on a negative curvature surface (left figures).
The folds in the red rectangles are equivalent to removing eye-like
areas as shown in the upper right figure. The buckled shape (the lower
right figure; the leaf-like object therein is the removed material)
of a flat disk due to an isolated fold is obtained by ``closing the
eye''. \label{fig:eye}}
\end{figure}

In addition to wrinkles, localized folds can also change the metric
\cite{diamant2011compression}. By attaching a paper disk to a negative
curvature surface one finds typical branch-like fold patterns as shown 
in Fig.~\ref{fig:eye} (the left two photos are the same deformed paper
disk from different perspectives). A light beam directly illuminates
the paper disk from above, so the folds are seen clearly as black
curves. The folds on the sheet can be roughly classified as principal
ones (in red rectangles) and fine ones (barely seen above the red
rectangle in the left lower figure). Their role in ``screening''
curvature is similar to topological defects in crystalline
order on a curved surface~\cite{bowick2009two}. The
feature of the observed folds is that the amount of the folded material
decreases towards their ends. Similar folds are also found in the
interior side of a bent tube where the curvature is negative \cite{houliara2010stability}.
We analyze an isolated fold to illustrate that it is compatible with
negative curvature geometry, which is analogous to the compatibility
of a seven-fold disclination with negative curvature geometry. The
effect of such folds is to remove an eye-like area from a flat disk.
The buckled shape (see the lower right figure in Fig.~\ref{fig:eye})
due to the fold, or equivalently the removal of an eye-like area,
is obtained by ``closing the eye''. The curvature of the buckled
shape is negative; the disk curves up along the fold and curves down
along the orthogonal direction. The removal of an eye-like area is
equivalent to inserting a wedge, because more material is removed
at the center than on the edge. The profile of the eye-like fold can
be determined by the curvature of the background geometry. The perimeter
of a circle with geodesic radius $r$ on a disk with an eye-like area
removed is estimated as $C(r)\approx2\pi r+4\left(h\left(0\right)-h\left(r\right)\right)$,
where $2h\left(r\right)$ is the width of a fold. For a small-sized
fold (in comparison with $\sqrt{1/\left|K_{G}\right|}$), one finds $h\left(r\right)=h\left(0\right)+\frac{\pi}{12}K_{G}r^{3}$
by inserting $C\left(r\right)$ into the expression for the Gaussian
curvature. From $h\left(r=L/2\right)=0$, we see that the length of
the fold is controlled by the curvature according to $L=2\left(\frac{12h\left(0\right)}{\pi\left|K_{G}\right|}\right)^{1/3}$.
In contrast, it is interesting to note that reversed eye-like folds are found in one's palms. It seems
that the two main lines - head line and heart line following the terminology
of palmistry - are compatible with positive curvatures as the width
of these lines increases from the center to the edge of a palm.

\begin{figure}
\centering{}\includegraphics[width=0.98\columnwidth]{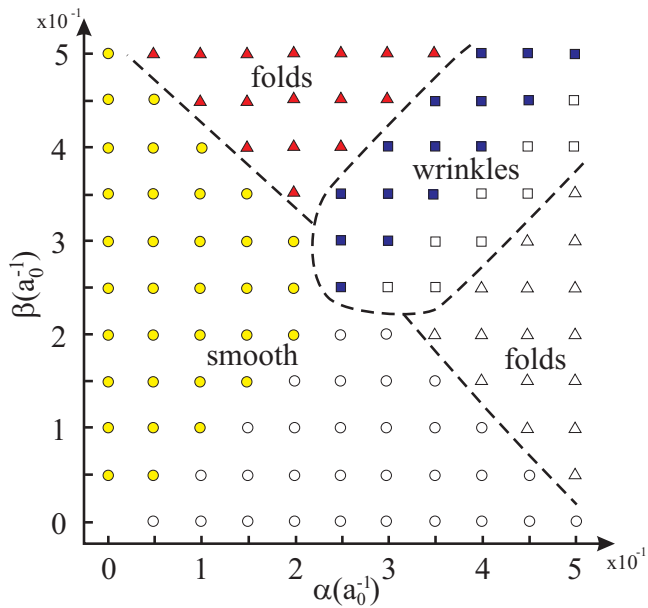}
\caption{(color online) The phase diagram of a deformed sheet on a negative
curvature surface, $z\left(x,y\right)=\alpha x^{2}-\beta y^{2}$.
Wrinkles (squares) on the sheet occur near the isotropic region ($\alpha/\beta\approx1$)
while folds (triangles) are found in the highly anisotropic regions
($\alpha/\beta$ far from unity). For very small $\alpha$ and $\beta$
the sheet is compliant with the adhering surface. This region is designated
as smooth (circles). Filled symbols indicate parameters for which
simulations were performed. The phase diagram is symmetric with respect
to the $\alpha=\beta$ line. Mirror images of the simulated points
are represented as open symbols. \label{fig:phase_diagram}}
\end{figure}

We now examine the transition between wrinkle and fold
patterns. In the regime of large surface tension, the ground state
of an elasto-capillary system is dominated by the surface energy difference
before and after a planar sheet is attached to the liquid interface:
$\Delta F=\sigma_{LA}A_{sheet}+\sigma_{LS}A_{coverage}-\sigma_{LA}A_{substrate}=-\sigma_{LS}A_{coverage}+\textrm{const}$.
$\sigma_{IJ}$ is the surface tension between phases $I$ and $J$
with $L$, $A$, and $S$ standing for liquid, air, and elastic sheet,
respectively. $A_{coverage}$ is the area of the liquid substrate
occupied by the elastic sheet, which is smaller than the area of the
sheet $A_{sheet}$ due to its deformation. $A_{substrate}$ is the
sum of the occupied and unoccupied substrate areas, which is a constant.
The surface energy turns out to depend only on $A_{coverage}$; the
larger it is, the smaller the energy is. Unlike a planar sheet on
a positive curvature surface, the deformation at the edge of a planar
sheet on a negative curvature surface may be ignored. Therefore, the
optimal contour shape of a deformed sheet on the liquid substrate
is the one that maximizes the coverage area while keeping the perimeter
fixed. This is exactly the classical isoperimetric problem, in this case on a negative
curvature surface. For constant curvature surfaces, the
classical isoperimetric solution in the Euclidean plane is also valid
with the circle in $\mathbb{E}^{2}$ being replaced by a geodesic
circle \cite{schmidt1939ueber}. On a general surface with varying
negative curvature, there is no exact mathematical result available.
The physical picture, however, is rather interesting: a deformed sheet
fully attached to a curved liquid substrate will migrate to the region
where it can extend as far as possible to maximize the contact area;
the driving forces are the capillary force and the release of the
bending energy in this curvature-driven migration process. 

In order to further explore the deformation patterns we have performed
numerical simulations of a thin elastic sheet adhered to a substrate
with negative curvature. For an isotropic material the stretching
and bending energies are given as \cite{koiter1966nonlinear} 
\begin{eqnarray}
E_{s} & = & \frac{t}{2}\int dA\frac{E}{1+\nu}\left(\frac{\nu}{1-\nu}u_{\alpha}^{\alpha}u_{\beta}^{\beta}+u_{\alpha}^{\beta}u_{\beta}^{\alpha}\right),\label{eq:stretch}\\
E_{b} & = & \frac{t^{3}}{24}\int dA\frac{E}{1+\nu}\left(\frac{\nu}{1-\nu}b_{\alpha}^{\alpha}b_{\beta}^{\beta}+b_{\alpha}^{\beta}b_{\beta}^{\alpha}\right).\label{eq:bend}
\end{eqnarray}
Here $\alpha,\beta\in\left\{ x,y\right\} $, $E$ is the three-dimensional Young's modulus and $Y=Et$ with $t$ being the sheet thickness; $u_{\alpha\beta}=\frac{1}{2}\left(g_{\alpha\beta}-\overline{g}_{\alpha\beta}\right)$
is the strain tensor with $\overline{g}_{\alpha\beta}$($g_{\alpha\beta}$)
being the metric tensor of the reference (deformed) state. The reference
metric is assumed to be flat, i.e., $\overline{g}_{\alpha\beta}=\delta_{\alpha\beta}$.
$b_{\alpha\beta}$ is the second fundamental form \cite{do1976differential},
$dA=\sqrt{\left|\bar{g}\right|}dxdy$ is the area element and $\nu$ is the Poisson's ratio. 
Finally, $u_{\alpha}^{\beta}=\overline{g}^{\beta\gamma}u_{\alpha\gamma}$
and $b_{\alpha}^{\beta}=\overline{g}^{\beta\gamma}b_{\alpha\gamma}$.
With the mean curvature $H\equiv\frac{1}{2}b_{\alpha}^{\alpha}$ and
the Gaussian curvature $K_{G}\equiv\det\left(b_{\alpha}^{\beta}\right)$,
$E_{b}=\int dA\kappa\left(2H^{2}-\left(1-\nu\right)K_{G}\right)$,
where the bending rigidity is $\kappa=\frac{t^{3}}{12}\frac{E}{\left(1-\nu^{2}\right)}$.
Since the deformations of the sheet do not change its topology, the
Gauss-Bonnet theorem \cite{do1976differential} ensures that $\int dAK_{G}$
is constant and the Gaussian curvature term can be omitted.

We find the relaxed shapes of the sheet by performing simulated annealing
Monte Carlo (MC) simulations of a discrete triangular mesh. The discrete
form of the stretching energy Eq. \eqref{eq:stretch} is \cite{parrinello1981polymorphic,sknepnek2012nonlinear}
\begin{equation}
E_{s}^{dis.}=\frac{Y}{8\left(1+\nu\right)}\sum_{T}\left(\frac{\nu}{1-\nu}\left(\mathrm{Tr}\hat{F}\right)^{2}+\mathrm{Tr}\hat{F}^{2}\right)A_{T},\label{eq:stretch_discrete}
\end{equation}
where $\hat{F}=\hat{\overline{g}}^{-1}\hat{g}-\hat{I}$, $A_{T}$
is a mesh triangle area, and the sum is carried out over all triangles.
$\hat{\overline{g}}$ and $\hat{g}$ are the discrete counterparts
of the reference and actual metric tensors, respectively, whose elements
are the scalar products of the two vectors spanning each triangle
before and after the deformation. By assuming that the sheet has no spontaneous
curvature,
the bending energy (Eq. \eqref{eq:bend}) of the discrete mesh is
\cite{brakke1978motion,brakke1992surface} 
\begin{equation}
E_{b}^{dis.}=2\kappa\sum_{i}\left(\frac{\left(\nabla_{\mathbf{r}_{i}}A_{i}\right)\cdot\mathbf{N}_{i}}{\mathbf{N}_{i}\cdot\mathbf{N}_{i}}\right)^{2}A_{i},\label{eq:bending_discrete}
\end{equation}
where $\nabla_{\mathbf{r}_{i}}$ is the gradient with respect to the
position of vertex $i$ and the sum is over all vertices.
$A_{i}=\frac{1}{3}\sum_{T\in\Omega_{i}}A_{T}$ is the vertex area,
with $\Omega_{i}$ being the vertex ``star'', i.e. the set of all
triangles that share vertex $i$. $A_{T}=\frac{1}{2}\left|\left(\mathbf{r}_{j}^{\left(T\right)}-\mathbf{r}_{i}\right)\times\left(\mathbf{r}_{k}^{\left(T\right)}-\mathbf{r}_{i}\right)\right|$,
where $\mathbf{r}_{i}$, $\mathbf{r}_{j}^{\left(T\right)}$ and $\mathbf{r}_{k}^{\left(T\right)}$
are the coordinate vectors of the vertices of $T$. We omitted $\left(T\right)$
in the superscript of vector $\mathbf{r}_{i}$ since it is shared
by all $T\in\Omega_{i}$. $\mathbf{N}_{i}=\nabla_{\mathbf{r}_{i}}V_{i}$
is a volume gradient, with the volume \cite{Siber06} $V_{i}=\frac{1}{6}\left|\mathbf{r}_{i}\cdot\sum_{T}\left(\mathbf{r}_{j}^{\left(T\right)}\times\mathbf{r}_{k}^{\left(T\right)}\right)\right|$.
The sheet is assumed to adhere to the substrate via a potential, $E_{adh.}=\frac{1}{2}\gamma\sum_{i}d_{i}^{2}A_{i}$,
where $d_{i}$ is the shortest Euclidean distance between the vertex
$i$ and the adhering surface.

The surface mesh with $\approx2 \times 10^{4}$ triangles was generated
by constructing a Delaunay triangulation \cite{cgal} of $\approx10^{4}$
randomly but evenly distributed points on a disk of radius $R\approx100a_{0}$,
where $a_{0}$ is the average distance between two neighboring points.
The initial state for each simulation was constructed by deforming the
planar mesh to perfectly comply with a prescribed shape of the substrate,
modeled as a hyperbolic paraboloid and parametrized as $z\left(x,y\right)=\alpha x^{2}-\beta y^{2}$.
A deformation pattern typically emerged within the first $5\cdot10^{4}$
MC sweeps and was further relaxed for an additional $2.5\times 10^{5}$
sweeps. In all simulations the energy scale is set by the bending
rigidity $\kappa$ and we set $Y=10^{3}\kappa/a_{0}^{2}$ (corresponding
to the thickness $t\approx0.03a_{0}$), $\gamma=10\kappa/a_{0}^{2}$,
and $\nu=1/3$.

\begin{figure}
\begin{centering}
\includegraphics[width=0.98\columnwidth]{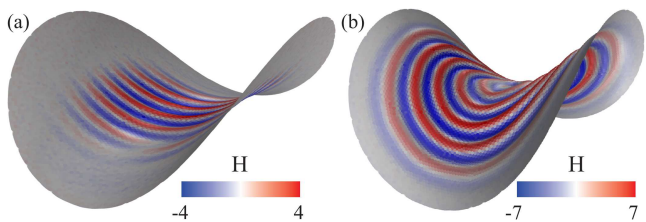}
\par\end{centering}

\caption{(color online) Snapshots of the relaxed conformation of a sheet adhered
to a negative curvature substrate modeled as a hyperbolic paraboloid
surface, $z\left(x,y\right)=\alpha x^{2}-\beta y^{2}$. For a substrate
with a highly anisotropic shape ($\alpha=0.1a_{0}^{-1}$, $\beta=0.5a_{0}^{-1}$)
one observes folds (a); sheets adhered to an isotripic substrate ($\alpha=\beta=0.5a_{0}^{-1}$)
develop wrinkles (b). Mean curvature, $H$, is measured in units of
$a_{0}^{-1}$. \label{fig:surface_snapshot}}
\end{figure}

In Fig.~\ref{fig:phase_diagram} we show a phase diagram of the wrinkle
and fold patterns for $\alpha,\beta\in\left[0,\dots,0.5\right]$,
measured in units of $a_{0}^{-1}$. We note that the Gaussian curvature
of a hyperbolic paraboloid surface is $K_{G}\left(x,y\right)=-4\alpha\beta/\left(1+4x^{2}\alpha^{2}+4y^{2}\beta^{2}\right)^{2}$
and has a maximally negative $K_{G}^{max}=-4\alpha\beta$ at $x=y=0$. For small
values of $\alpha,\beta\lesssim0.2a_{0}^{-1}$ the Gaussian curvature
is very small, $\left|K_{G}\right|\lesssim0.15a_{0}^{-2}$ and the
strong adhesion forces prevent the sheet from deforming; instead
it remains smooth and perfectly compliant to the substrate. As $\alpha$ and
$\beta$ increase two distinct deformed states form, folds and wrinkles.
These can be distinguished by the isometry condition as follows. The
inextensibility of a sheet sets an upper limit $2R$, the diameter
of the planar sheet, as the maximal geodesic distance between two
arbitrary points on a sheet contour. If the determined sheet contour
has the maximum geodesic diameter equal to (smaller than) $2R$, then
the sheet is recognized as having folds (wrinkles). An isotropic
Gaussian curvature ($\alpha=\beta$, since $K_{G}=f\left(\alpha^2 x^{2}+\beta^2 y^{2}\right)$
\cite{struik1988lectures}) imposes either isotropic tangential stretching
or isotropic radial compression on the sheet, which is expected to
result in wrinkles. As the Gaussian curvature grows increasingly anisotropic
(the ratio $\alpha/\beta$ deviates from unity), the imposed anisotropic
stretching and compression on the sheet are expected to generate folds
which are themselves anisotropic objects. This shows that wrinkles
occur near the isotropic region while folds arise in the highly anisotropic
regions. A typical configuration with folds is shown in Fig.~\ref{fig:surface_snapshot}a,
while a typical wrinkle pattern is shown in Fig.~\ref{fig:surface_snapshot}b.
Finally, in the highly anisotropic case, $\alpha\ll\beta$,
$K_{G}\approx0$ and the substrate is nearly cylindrical; in this case the deformation
of the sheet is nearly isometric and no wrinkles or folds form. 

In conclusion, using geometric arguments and numerical simulations
of a thin-sheet elastic model we study the curvature-driven wrinkle/fold
patterns of a flat sheet adhered to a negative curvature substrate.
We analyze two types of structure, concentric wrinkles and eye-like
folds, that are compatible with negative curvature
liquid substrates and discuss the transition between these two states
driven by the anisotropy of the background geometry. The ability of
wrinkles/folds to deform a plane to a curved surface may find potential
applications. Consider a flat sheet with a pre-designed wrinkle/fold
pattern like the lines on a palm. By controlling the on/off status of
the wrinkles/folds, a planar sheet can be programmed to buckle to
a desired shape. This may lead to potential applications in maximizing
sunshine harvest by designing the shape of ultra thin flexible
solar cells \cite{kaltenbrunner2012ultrathin}. Our study also sheds
light on the reverse problem of attaching a curved shell to a flat
substrate, e.g., the adhesion of a cell on a flat substrate \cite{sackmann2002cell}. 

We thank Jennifer Schwarz for kindly allowing us to use her computer
resources. This work was supported by the National Science Foundation
grant DMR-0808812 and by funds from Syracuse University. 

\bibliographystyle{apsrev}

\bibliography{references,elasticity}

\end{document}